\documentclass[english,reprint, aps]{revtex4-1}
\usepackage[LGR,T1]{fontenc}
\usepackage[latin9]{inputenc}
\usepackage{textcomp}
\usepackage{amsmath}
\usepackage{amsthm}
\usepackage{graphicx}
\usepackage{subscript}

\makeatletter

\DeclareRobustCommand{\greektext}{%
  \fontencoding{LGR}\selectfont\def\encodingdefault{LGR}}
\DeclareRobustCommand{\textgreek}[1]{\leavevmode{\greektext #1}}
\ProvideTextCommand{\~}{LGR}[1]{\char126#1}

\providecommand{\tabularnewline}{\\}

\numberwithin{equation}{section}
\numberwithin{figure}{section}

\makeatother

\usepackage{babel}
\begin{document}
\title{New Born-Oppenheimer molecular dynamics based on the extended Hückel
method: first results and future developments}
\author{Pedro E. M. Lopes}
\affiliation{Rua Almirante Reis, Nº 28A, 2º Esq, 2330-099 Entroncamento, Portugal }
\altaffiliation{11 Warren Lodge Ct 1D, Cockeysville, MD 21030, USA}

\email{plopesuk@yahoo.co.uk }

\homepage{www.fastcompchem.com}

\selectlanguage{english}%
\begin{abstract}
Computational chemistry at the atomic level has largely branched into
two major fields, one based on quantum mechanics and the other on
molecular mechanics using classical force fields. Because of high
computational costs, quantum mechanical methods have been typically
relegated to the study of small systems. Classical force field methods
can describe systems with millions of atoms, but suffer from well
known problems. For example, these methods have problems describing
the rich coordination chemistry of transition metals or physical phenomena
such as charge transfer. The requirement of specific parametrization
also limits their applicability. There is clearly a need to develop
new computational methods based on quantum mechanics to study large
and heterogeneous systems. Quantum based methods are typically limited
by the calculation of two-electron integrals and diagonalization of
large matrices. Our initial work focused on the development of fast
techniques for the calculation of two-electron integrals. In this
publication the diagonalization problem is addressed and results from
molecular dynamics simulations of alanine decamer in gas-phase using
a new fast pseudo-diagonalization method are presented. The Hamiltonian
is based on the standard Extended Hückel approach, supplemented with
a term to correct electrostatic interactions. Besides presenting results
from the new algorithm, this publication also lays the requirements
for a new quantum mechanical method and introduces the extended Hückel
method as a viable base to be developed in the future.
\end{abstract}
\maketitle

\section{Introduction}

The development of atomistic computational methods for chemistry,
biology and physics has a long and notable tradition, having evolved
from pure academic curiosities to indispensable tools of great economic
value. Notable examples where computational methods take a central
role include the development of new drugs \citep{Macalino2015,Riccardi2018}
and the design of new enzymes and catalysts \citep{Kries2013,Hilvert2013}.
Innumerable methods have been developed spanning very different approximations
and targeting different systems and systems sizes. They can typically
be grouped into certain levels, each one corresponding to a certain
complexity and functionality. Each level is typically inter-related
to the levels below and contributes to the levels above. At the lowest
level, there is Quantum Mechanics (QM), with progress in algorithms
and computer hardware making possible to model systems with a few
thousand atoms \citep{Ratcliff2017}, although simulations of systems
with millions of atoms have been reported \citep{Hoshi2013}. Those
calculations required sophisticated super-computers with tens of thousands
of processors and are not generally accessible to most academic and
industry users. Highly accurate calculations are also possible for
up to a few dozen atoms, for example using coupled-cluster techniques
\citep{Bartlett2011}. The next level of simulation methods can handle
systems with up to millions of atoms and are based on classical Newtonian
mechanics and empirical Force Fields (FFs) \citep{Zhao2013}. At the
upper end, the mesoscale describes systems of billions of atoms using
very approximate methods that still reflect molecular effects. The
level of accuracy of each computational methodology deceases significantly
on moving from the quantum level to the mesoscale. This work reports
the initial steps in the development of new methodologies for fast
and accurate methods at the quantum level for large systems on commodity
hardware. Next, the need for atomistic methods based on QM is described,
due to known problems of classical FFs. Then the Extended Hückel (EH)
method is briefly described since it is a good base for the new computational
method. Finally, results from gas-phase molecular dynamics simulations
of alanine decamer, (Ala)\textsubscript{10}, are presented and discussed.

\section{\label{sec:Need new methods 2}Need of new computational methods
based on quantum mechanics }

Quantum Mechanical methods are in general generic and applicable to
most problems in chemistry and biology. The major problem of QM based
methods is their limitation to small systems, which typically excludes
most of the large systems in biology, physics and chemistry. FF methods,
on the other hand, were designed to be fast and applicable to very
large systems. In 2006, for example, the first molecular dynamics
simulation of a system with more than a million atoms was reported
\citep{Freddolino2006}. FFs abandoned the QM principles and the energy
function is purely classical. The harmonic nature of empirical FFs
does not allow the study of systems with breaking and formation of
chemical bonds and since FFs are parametrized for certain classes
of compounds their applicability is limited. Besides reactivity studies,
classical FFs have other limitations that impart their effectiveness.
One notorious problem is the difficulty of FFs in describing systems
with transition metals. This is a very significant area since according
to some sources approximately half of all proteins are metalloproteins
\citep{Thomson1998}, meaning that approximately half of all proteins
cannot, or are incorrectly studied, with classical FFs. Another limitation
of classical FFs is the description of polarization and Charge-Transfer
(CT) effects. Methodologies to describe polarization effects have
been incorporated into classical FFs with notable examples being AMOEBA
\citep{Ponder2010,Shi2013} and the Drude polarizable Force Field
(polFF) \citep{Huang2014,Lopes2015,Baker2015,Huang2018}. Charge-transfer,
being a pure quantum effect, is considerably more difficult to describe
with classical methods. It remains a severe issue with empirical FFs.
QM studies have shown that CT effects account for approximately one-third
of the binding energy in a neutral water dimer \citep{Khaliullin2009},
and a similar amount (22-35\% depending on the semi-empirical method)
in protein-protein interactions \citep{Ababou2007}. \textit{Ab initio}
molecular dynamics studies of BPTI in water and vacuum also revealed
significant CT between the solvent and the protein. Furthermore, for
the simulations in vacuum a very significant intra-molecular CT was
found between the neutral and charged residues. Interestingly, upon
solvation the formally neutral residues remained neutral \citep{Ufimtsev2011}.
The reliance of classical FFs on parameterization has positive and
negative aspects. On the positive side, classical FFs can be made
very accurate in reproducing properties in gas and condensed phases.
For example, classical water models \citep{Yu2013,Demerdash2018}
are able to reproduce the structure of liquid water considerably better
than pure Density Functional Theory (DFT) \citep{Kuo2004,Wang2011}.
On the negative side, development of classical FFs is a painstaking
process that includes multiple fittings and careful judgment to obtain
the best compromise in reproducing the experimental and \textit{ab
initio} target data \citep{Foloppe2000}. The problem is exacerbated
with polFFs where transferability of the electrostatic parameters
can be lost due to cooperative effects, as was the case with the Drude
polFF (see references \citep{Lopes2015} and \citep{Lopes2013} for
an example). The ideal parameterization scheme for polFFs would require
reliable reference data for large systems, in gas and condensed phase,
from large scale QM calculations, that is difficult to compute using
current methods. Development of more accurate polFFs will clearly
benefit from availability of fast and accurate large scale QM methods.
There have been many attempts to remedy the deficiencies of classical
FFs. Illustrative examples are the development of reactive FFs and
development of theories to quantify effects of d-electrons. Reactive
FFs have been developed to describe chemical bonding without expensive
QM calculations, thus allowing studies of reactive events \citep{Senftle2016}.
Deeth and co-workers added energy terms from d-electrons derived from
ligand-field theory to classical FFs to study systems with transition
metal complexes \citep{Deeth2009}. These efforts, despite improving
traditional classical FFs are not effective replacements for a full
QM description. From the discussion above it is clear that there is
a great and urgent need to develop fast and accurate QM methods capable
of studying large and complex systems across multiple scientific areas.
Although such a computational methodology can be understood as a direct
competitor to classical FFs, it is better to see it as another element
of the multiscale ladder with potential close integration with classical
FFs. In our vision, the new QM method will be reserved to study systems,
or parts of systems, where quantum effects are preponderant, leaving
the remaining parts for classical FFs. Because studies of very large
systems will be possible, the new QM methods will also be used in
the parameterization of very accurate polFFs. The quantum and classical
approaches will also be interfaced together in ways reminiscent of
QM/MM methods.

\section{\label{sec:A vision for the future}A vision for the future built
step by step}

The problems limiting the development of fast QM methods for large
systems are well known. Calculation of two-electron integrals scales
formally as N\textsuperscript{4}, due to the four different functions
in the integral. Several techniques that include integral screenings
and fittings to decrease the dimensionality of the integral have been
proposed (see \citep{Reine2012} for appropriate references), but
computation of two-electron integrals remains a formidable task. Another
limitation of typical QM methods is diagonalization of the Hamiltonian,
which has a N\textsuperscript{3} scaling. Much effort has been put
into developing alternatives to diagonalization. For electronic structure
calculations several alternatives to formal diagonalization have been
developed including partition of a larger problem into smaller more
amenable problems \citep{Lee1998,Dixon2002} and direct minimization
of the density matrix \citep{Arita2014,Mniszewski2015}. Development
of novel QM algorithms for physics, chemistry and biology has to,
invariably, address both problems. Recently, we developed a fast algorithm
to compute two-electron integrals by approximation, using nested bi-
and single-dimensional Chebyshev polynomials \citep{Lopes2017}. This
algorithm is not a direct replacement to standard methodologies, since
it is limited to a fixed basis due to parameterization. It is, however,
very fast and for the basis set of choice, allowing development of
Hartree-Fock (HF) or DFT methods that are able to describe very large
systems. A new methodology for pseudo-diagonalization has also been
recently developed, and the first results are show in this publication
(see below). A full description of the methodology will be presented
in a dedicated publication. When both of our most recent works are
combined, they will allow development of a new generation of fast
and accurate QM methods for large and complex systems. The goal is
to develop a fast and accurate QM method that is capable of studying
the dynamics of systems with 10-100,000 atoms on commodity hardware.
Ideally, the new computational method needs to have several distinct
features: 
\begin{itemize}
\item Be applicable to a wider range of large systems, over greater time
scales, including gas phase, liquid and solid state. 
\item Be flexible and rely on standard optimized libraries, for example
optimized implementations of Blas and Lapack. This means the method
will benefit from high level optimizations without incurring further
development costs, and will be fully inter-operable between various
computing platforms and operating systems. 
\item Be affordable and capable of running on workstations or lower-cost
parallel systems, not only expensive supercomputers. For this purpose
it has to take advantage of new computing paradigms such as Graphical
Processing Units (GPUs). 
\item Be able to reproduce experimental gas and condensed phase properties.
This is, perhaps, the most important factor in the success of classical
FFs and the same concepts need to be incorporated in the development
of the new method. To my best knowledge, this is the first time that
this important concept for classical FF development is being used
in QM methods development.
\end{itemize}

\section{\label{sec:highlights EH}Highlights of the extended-Hückel method.
Possible basis for future computational method }

The desirable features for the new QM computational method outlined
above find a good match on existing tight-binding approaches, such
as the EH. The EH method is conceptually very simple, which tend to
yield faster algorithms, and has been connected to the standard HF
algorithm, showing a clear path for optimization and enhanced accuracy.
It is very interesting that in two very different periods, spanning
a period of fifty years, the connection between EH and HF has been
established. Blyholder and Coulson \citep{Blyholder1968} made the
connection using arguments based on the Mulliken approximation of
two-electron integrals. Recently, Akimov and Prezhdo \citep{Akimov2015}
made the connection between HF and Self-Consistent Extended Hückel
(SC-EH). Because of its simplicity it is also a great tool to use
in the classroom \citep{Arita2014}. The method has been applied to
many diverse systems, including studies of relativistic effects \citep{Pyykko1981}.
Although it had been initially applied mostly in studies of organic
compounds, it was later extended to study inorganic and organometallic
complexes \citep{Hoffmann1981}, including periodic systems and nanoscale
materials \citep{Hoffmann1988,Nishino2013}. It was also used in tight-binding
calculations of molecular excited states \citep{Rincon2008} and it
has also been reformulated to include unrestricted calculations \citep{Kitamura2000}.
The initial formulation was non-iterative with the matrix elements
of the Hamiltonian, $H_{ij}$ being charge-independent and computed
as

\begin{equation}
H_{ij}=K_{ij}\frac{\left(\epsilon_{i}+\epsilon_{j}\right)}{2}S_{ij}\label{eq: equacao simples}
\end{equation}

\bigskip{}

where $S_{ij}$ are the overlap integrals between Atomic Orbitals
(AO) and $\varepsilon_{i}$ are orbital energies. AOs were typically
described by Slater type functions (STF). The parameter $K_{ij}$
is fixed in this simple formulation and Hoffmann suggested the value
of 1.75 \citep{Hoffmann1963}. 

The simplest EH Hamiltonian of Eq. \ref{eq: equacao simples} was
corrected several times addressing mainly two aspects: (1) modification
of the electronic Hamiltonian, often through different formulations
of the parameter $K_{ij}$, and (2) addition of nuclear repulsion
and nucleus-electron attraction terms. Several formulas were suggested
to correct the Hamiltonian including parameterizations by Cusachs
\citep{Cusachs1965}, Kalman \citep{Kalman1973}, Anderson \citep{Anderson1975},
and Ammeter et al. \citep{Ammeter1978}. The latest parameterization
of the Hamiltonian due to Calzaferri \textit{et al.} \citep{Calzaferri1989,Brandle1993,Brandle1993a,Calzaferri1995}
introduced a distance dependent formula for $K_{ij}$, in order to
make it larger than 1 for intermediate and large inter-atomic separations, 

\begin{equation}
K_{ij}=\left\{ 1+\kappa_{ij}\exp\left(-\delta\left(R_{ij}-d_{0}\right)\right)\right\} \label{eq: equacao Hij dependente da distancia}
\end{equation}

with $\kappa_{ij}$ and $\delta$ being positive parameters, and $d_{0}$
a parameter defined by Calzaferri as the sum of orbital radii. In
reference \citep{Calzaferri1989} formulas for $\kappa_{ij}$ and
$d_{0}$ are given.

Additional nuclear repulsion and nucleus-electron attraction corrections
were added initially by Anderson and Hoffmann \citep{Anderson1974}.
Carbó \textit{et al.} \citep{Carbo1977} also introduced electrostatic
corrections to the EH method. Lastly, Calzaferri and co-workers \citep{Calzaferri1989,Brandle1993,Brandle1993a,Calzaferri1995}
defined the electrostatic terms as 

\begin{widetext}

\begin{equation}
E_{corr}\left(\overrightarrow{R}\right)=\sum_{\begin{array}{c}
A,B\\
A<B
\end{array}}\left\{ \frac{Z_{A}Z_{B}}{R_{AB}}-\frac{1}{2}\left[Z_{A}\int\frac{\rho_{B}\left(\overrightarrow{r}\right)}{|\overrightarrow{R}_{AB}-\overrightarrow{r}|}dr+Z_{B}\int\frac{\rho_{A}\left(\overrightarrow{r}\right)}{|\overrightarrow{R}_{AB}-\overrightarrow{r}|}dr\right]\right\} \label{eq: equacao correccoes electrostaticas}
\end{equation}

with the integrals being computed as

\begin{equation}
\int\frac{\rho\left(\overrightarrow{r}\right)}{|\overrightarrow{R}-\overrightarrow{r}|}dr=\frac{1}{R}\sum_{n,l}b_{n,l}\left[1-\frac{\exp\left(-2\zeta_{n,l}R\right)}{n,l}\sum_{p=1}^{2n}\left(2\zeta_{n,l}R\right)^{2n-p}\frac{p}{\left(2n-p\right)!}\right]\label{eq: equacao dos integrais}
\end{equation}

\end{widetext}

The coefficients $b_{n,l}$ are the occupation numbers of the corresponding
AOs with exponents $\zeta_{n,l}$, and $n,l$ are the principal and
the azimuthal quantum numbers. Because the electrostatic correction
term is a posteriori correction, added to the EH electronic energy,
the charge densities are determined non-self consistently by the EH
Hamiltonian. The advantage of this scheme is faster calculations since
no self-consistent field (SCF) cycles are needed, while still adding
important contributions to the total energy. It is noteworthy that
Eq. \ref{eq: equacao correccoes electrostaticas} does not include
any electron-electron repulsion terms. Our recent work on fast algorithms
for two-electron integrals \citep{Lopes2017} will allow explicit
accounting of electron-electron interactions. Self-consistent schemes
were also added to the EH formulations, in order to better describe
charge transfer effects \citep{Kalman1973,Mukhopadhyay1981}. The
resulting equations must be solved iteratively because of the dependence
of the Hamiltonian on the charge distributions and vice-versa. The
algorithms are very similar to SCC-DFTB \citep{Elstner2006}.

\begin{table}
\caption{\label{tab: valores optimizados}Optimized parameters of the atomic
orbitals in a single-$\zeta$ representation of Slater type functions
(STFs). $\varepsilon_{s}$ and $\varepsilon_{p}$ are the s and p
orbital energies and $\zeta_{s}$ and $\zeta_{p}$ are the corresponding
exponential parameters of the STFs}

\begin{tabular}{|c|c|c|c|c|}
\hline 
 & $\varepsilon_{s}$ & $\zeta_{s}$  & $\varepsilon_{p}$ & $\zeta_{p}$ \tabularnewline
\hline 
\hline 
H & -16.2866 & 1.2784 &  & \tabularnewline
\hline 
C & -25.6725  & 1.6958 & -13.6697 & 1.8884\tabularnewline
\hline 
N & -21.5711 & 1.8000 & -13.5185 & 2.5758\tabularnewline
\hline 
O & -25.8400 & 1.9867 & -13.5413 & 2.5942\tabularnewline
\hline 
\end{tabular}
\end{table}

\section{\label{sec:dinamica molecular}Testing the pseudo-diagonalization
method: Born-Oppenheimer molecular dynamics simulations }

\begin{figure}
\includegraphics[scale=0.25]{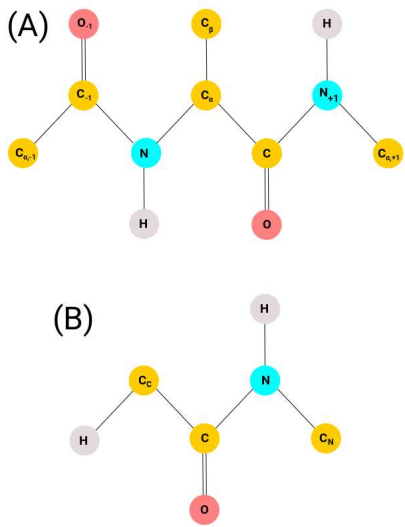}

\caption{\label{fig:Identification-of-relevant}Identification of relevant
atoms of (ALA)\protect\textsubscript{10} (A) and NMA (B) used in
the analysis}
\end{figure}
\begin{figure}
\includegraphics{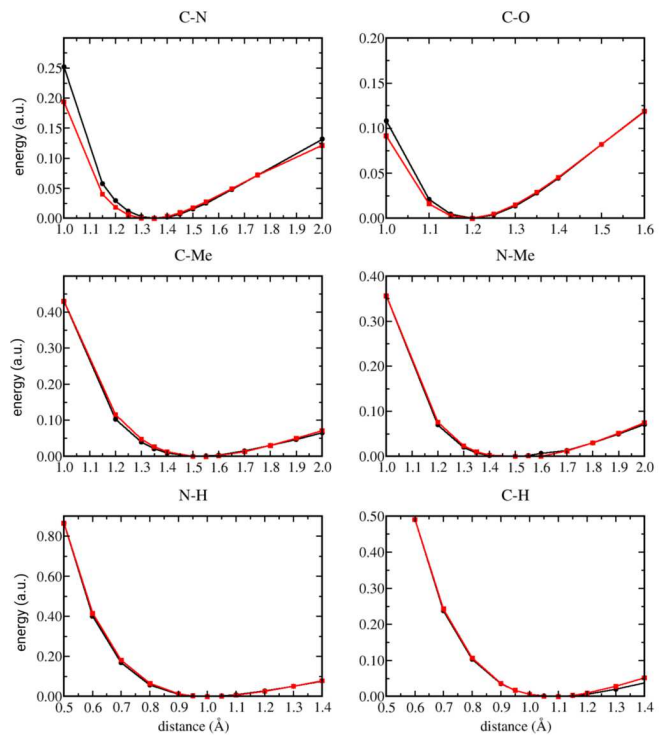}

\caption{\label{fig:Potential-energy-surfaces}Potential energy surfaces for
stretching of specific bonds of NMA. The reference values from HF/cc-pVDZ
calculations are in black and the fitted values using the EH Hamiltonian
in red}
\end{figure}
In order to demonstrate the capabilities of the newly developed algorithms
for fast density optimization and pseudo-diagonalization and the performance
of the basic EH Hamiltonian, Born-Oppenheimer molecular dynamics simulations
were performed. The test system is (Ala)\textsubscript{10} in gas
phase. The system was deliberately kept small to allow extensive monitoring
of the calculated energies relative to the true values obtained by
diagonalization along the trajectory. The Hamiltonian was based on
the EH approach with the standard Hamiltonian (Eq. \ref{eq: equacao simples})
with additional nuclear repulsion and nuclear-electronic attraction
terms described by Eq. \ref{eq: equacao correccoes electrostaticas}.
All calculations were performed on a single AMD Ryzen 1700X processor.
The initial goal is to establish the baseline performance of the method
without speed boosts due to multiprocessing and parallelization. The
scalability of the algorithms is an important feature of the development
process though, and all algorithms are being developed considering
parallelization using standard CPU and emerging computer architectures
such as GPUs. 

\begin{widetext}

\begin{table}
\caption{\label{tab: angulos e distancias}Selected bond distances and angles
of alanine decamer from the simulations and experiment}

\begin{tabular}{ccccccc}
\hline 
\multicolumn{7}{c}{Angles (degrees)}\tabularnewline
\hline 
 & C\textsubscript{\textgreek{b}}-C\textsubscript{\textgreek{a}}-C  & N-C\textsubscript{\textgreek{a}}-C\textsubscript{\textgreek{b}} & C\textsubscript{\textgreek{a}}-C-O & C\textsubscript{\textgreek{a}}-C-N\textsubscript{+1} & O-C-N\textsubscript{+1} & C\textsubscript{-1}-N-C\textsubscript{\textgreek{a}}\tabularnewline
MD  & 111.1\textpm 4.1 & 106.7\textpm 3.8 & 121.3\textpm 3.1 & 118.2\textpm 3.6 & 120.2\textpm 2.8 & 123.2\textpm 4.0\tabularnewline
Exp. \citep{Balasco2017}\textsuperscript{{*}} & 109.7\textpm 1.5  & 110.9\textpm 1.4 & 120.4\textpm 0.9 & 116.4\textpm 1.1 & 123.2\textpm 0.9 & 122.5\textpm 1.3\tabularnewline
\hline 
\multicolumn{7}{c}{Distances (Å)}\tabularnewline
\hline 
 & C\textsubscript{\textgreek{a}}-C  & N-C\textsubscript{\textgreek{a}} & C-N\textsubscript{+1}  & C-O & C\textsubscript{\textgreek{a}}-C\textsubscript{\textgreek{b}}  & \tabularnewline
MD & 1.503\textpm 0.021  & 1.479\textpm 0.021 & 1.338\textpm 0.027  & 1.199\textpm 0.012  & 1.618\textpm 0.022 & \tabularnewline
Exp. \citep{Improta2015}\textsuperscript{\#} & 1.525, 1.531 & 1.456, 1.461 & 1.332, 1.336 & 1.232, 1.235 & 1.525, 1.531  & \tabularnewline
HF/cc-pVDZ\textsuperscript{{*}{*}} & 1.514 & 1.447 & 1.352 & 1.199 & NA & \tabularnewline
\hline 
\multicolumn{7}{l}{{\footnotesize{}{*} Values of non-Gly/non-Pro residues in \textgreek{b}-structure}}\tabularnewline
\multicolumn{7}{l}{{\footnotesize{}\# Average values for different (\textgreek{y},\textgreek{f})
regions}}\tabularnewline
\multicolumn{7}{l}{{\footnotesize{}{*}{*} Optimized values for NMA. See Figure \ref{fig:Identification-of-relevant}
for naming of the atoms}}\tabularnewline
\end{tabular}
\end{table}
\end{widetext}

Initially, the values of each $\varepsilon_{i}$, Slater exponents
$\zeta_{i}$ and the $K_{ij}$ value were optimized based on fittings
to potential energy surfaces for bond stretching and shorting around
the equilibrium in N-methyl acetamide (NMA). NMA is the smallest molecule
prototyping a chemical bond and is used extensively in the development
of classical FFs \citep{Harder2008,Lin2013}. Six bond lengths were
fitted in total: C(sp\textsuperscript{2})-N, C-O, N-C(sp\textsuperscript{3}),
N-H, C(sp\textsuperscript{3})-H and C(sp\textsuperscript{2})-C(sp\textsuperscript{3}).
The C(sp\textsuperscript{3})-C(sp\textsuperscript{3}) bond in alanine
was deliberately not fitted to test transferability of the C parameters.
The target potential energy surfaces were obtained at the HF/cc-pVDZ
and restricted to the vicinity of the minimum. Since this work is
only illustrative of the capabilities of the new algorithm for pseudo-diagonalization
and of the EH method, there is no need for higher level ab initio
target data. The parameters were fitted freely using the same simulated
annealing procedure used to develop the Drude polFF \citep{Yu2013,Lopes2013}.
In Figure \ref{fig:Potential-energy-surfaces} the plots of the reference
and fitted potential energy surfaces are show (see the fitted parameters
in Table \ref{tab: valores optimizados}). For each plot the agreement
between the target and the computed energies is nearly perfect. This
adds to the great potential of the basic EH method to be a suitable
basis to develop a new class of methods for computational quantum
chemistry.

\begin{figure}
\includegraphics[scale=0.2]{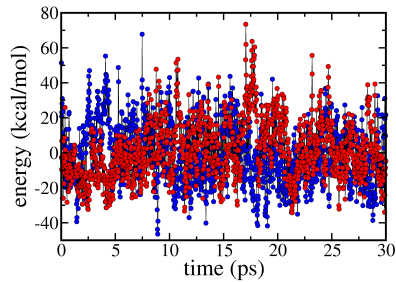}

\caption{\label{fig:Total-energy-fluctuation}Total energy fluctuation states
(bottom) for alanine decamer in gas phase at 300 K. The average total
energy is -17,598.5 kcal/mol. Red is simulation 1 and blue is simulation
2}
\end{figure}
\begin{figure}
\includegraphics[scale=0.2]{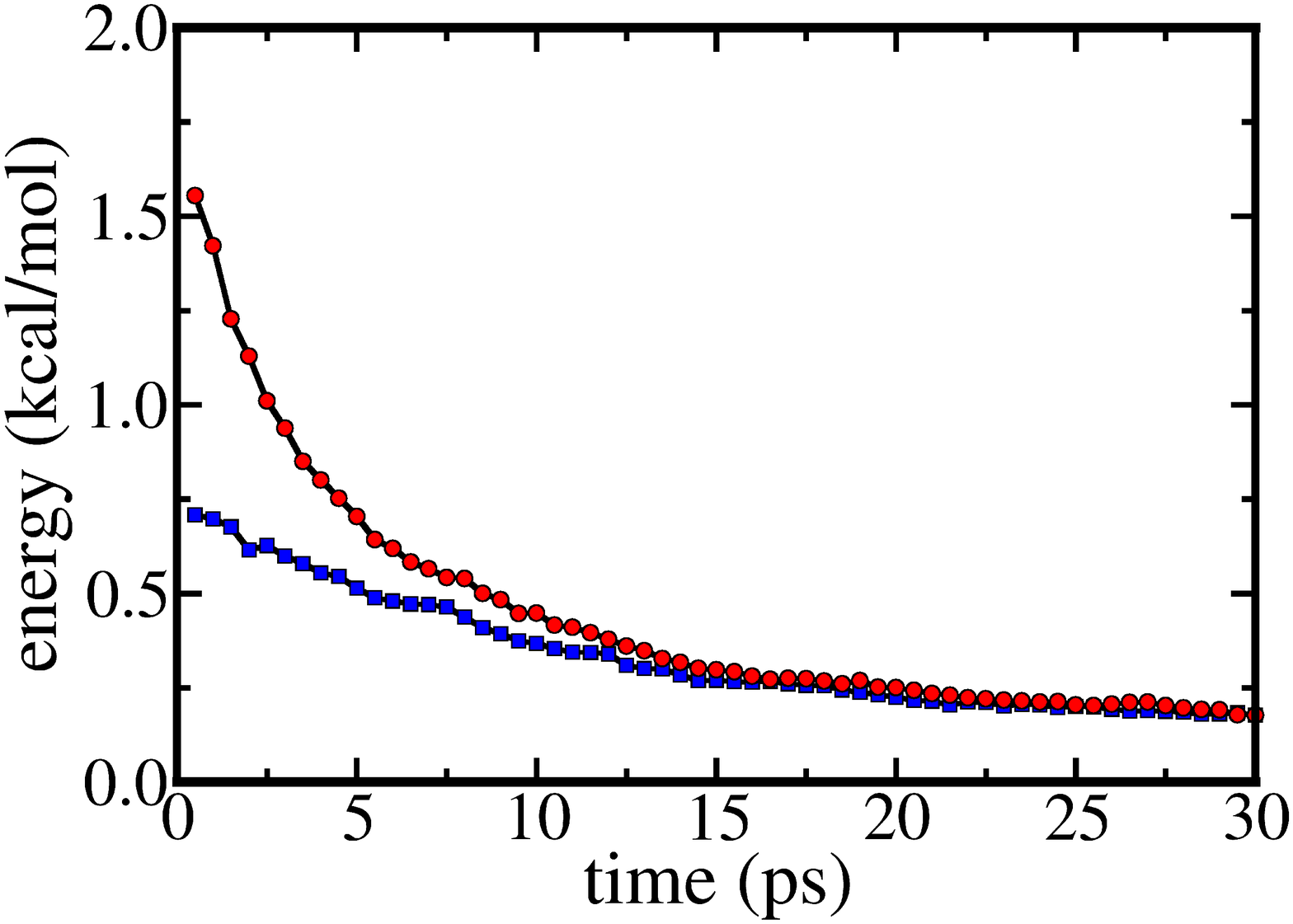}

\caption{\label{fig:Difference-between-potential-ener}Difference between the
potential energies from the new pseudo-diagonalization method and
the exact values from diagonalization. Red is simulation 1 and blue
is simulation 2}
\end{figure}
After suitable parameters have been developed, MD simulations were
performed. The test system was (ALA)\textsubscript{10} in gas phase.
These results come with an important disclaimer since the EH parameters
were not fitted taking into consideration the relative energies of
different conformers. Thus, there is no guarantee that sampling is
appropriate, for example as a function of the \textgreek{f} and \textgreek{y}
torsions. The Newton\textquoteright s equations of motion were integrated
using the velocity Verlet algorithm with 1 fs timestep and the temperature
(300 K) was maintained using the Berendsen thermostat \citep{Berendsen1984}.
In Figure \ref{fig:Total-energy-fluctuation} the total energy is
plotted for two simulations of 30 ps each. The second simulation (in
blue) started from the last frame of the first simulation (in red)
with randomized velocities. The energy is well conserved and there
is no apparent drift. It has been reported that with methods dependent
on self-consistent iterations significant drifts of the total energy
can occur due to incompleteness of optimization \citep{Herbert2005}.
Niklasson and co-workers have proposed the XL-BOMD method to remedy
this problem \citep{Aradi2015,Souvatzis2013,Steneteg2010}. In the
present case, the EH Hamiltonian is immune due to its non-self consistency. 

\begin{figure}
\includegraphics{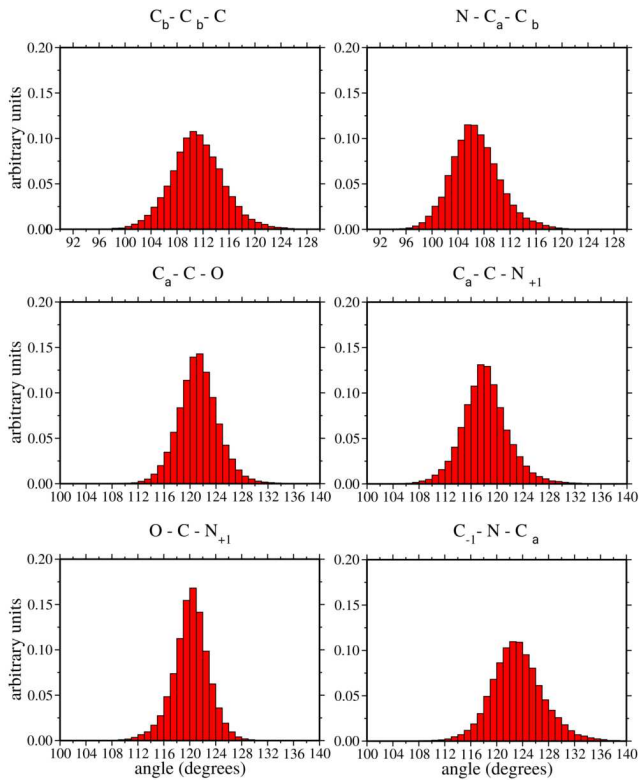}

\caption{\label{fig:Histograms-illustrating-the}Histograms illustrating the
distribution of selected bond angles during the MD simulations. For
nomenclature of the atoms refer to Figure \ref{fig:Identification-of-relevant}}
\end{figure}
Despite the important feature of the total energy not drifting, it
is important to understand how the potential energies (and the forces)
from the pseudo-diagonalization algorithm compare with the true energies
obtained from diagonalization. In Figure \ref{fig:Difference-between-potential-ener}
the differences of potential energies between the new pseudo-diagonalization
and true diagonalization are plotted. Remarkably, the energy difference
shows a continuous decrease for both simulations. At 30 ps the difference
is very small relative to the true values from diagonalization (0.18
kcal/mol in both cases for an average total energy of -17,598.50 kcal/mol),
which validates the new algorithm as a viable tool for electronic
structure calculations. In future works this behavior will be analyzed
thoroughly.

Next, sampling of selected bond distances, angles and torsions from
the simulations will be analyzed. The C\textsubscript{\textgreek{a}}-C,
N-C\textsubscript{\textgreek{a}}, C-N\textsubscript{+1}, C-O, C\textsubscript{\textgreek{a}}-C\textsubscript{\textgreek{b}}
bonds, C\textsubscript{\textgreek{b}}\textsubscript{}-C\textsubscript{\textgreek{a}}-C,
N-C\textsubscript{\textgreek{a}}-C\textsubscript{\textgreek{b}},
C\textsubscript{\textgreek{a}}-C-O, C\textsubscript{\textgreek{a}}-C-N\textsubscript{+1},
O-C-N\textsubscript{+1}, C\textsubscript{-1}-N-C\textsubscript{\textgreek{a}},
angles and O-C-C\textsubscript{\textgreek{a}}-N\textsubscript{+1},
H-N\textsubscript{+1}-C-C\textsubscript{\textgreek{a}+1} torsions
are considered for analysis (See Figure \ref{fig:Identification-of-relevant}(A)
for naming of the atoms). In Table \ref{tab: angulos e distancias}
the averaged values from the simulations, together with their experimental
and NMA QM equivalents are presented. Figure \ref{fig:Histograms-illustrating-the}
shows histograms for the distribution of bond angles from the simulations.
For all angles normal distributions are observed. In future publications,
the distribution of bond angles from BOMD will be compared with experimental
values from high-resolution crystal data. Starting with the bond distances,
it is apparent that the simple EH Hamiltonian reproduces outstandingly
well all bonds with the exception of the C\textsubscript{\textgreek{a}}-C\textsubscript{\textgreek{b}}
bond. For the C\textsubscript{\textgreek{a}}-C\textsubscript{\textgreek{b}}
bond the equilibrium value is \textasciitilde 1.53 Å whereas the
averaged value from the MD simulations is 1.618 Å. In the polypeptide
some distances increase, while others decrease, relative to the optimized
values from NMA. In future works the target geometries will be derived
from geometry optimizations at a higher level of theory with correlation.
The bond angles also remain very close to the experimental values.
The largest deviations are for N-C\textsubscript{\textgreek{a}}-C\textsubscript{\textgreek{b}}
and O-C-N\textsubscript{+1}, with 4.2 and 3.0°, respectively. It
is interesting to note that the largest deviation for the bond angles
also involves C\textsubscript{\textgreek{b}}. The planarity of the
peptide bonds is maintained along the simulation with the out-of-plane
torsions of 180.4° for O-C-C\textsubscript{\textgreek{a}}-N\textsubscript{+1}
and 179.7° for H-N\textsubscript{+1}-C-C\textsubscript{\textgreek{a}+1}.

\section{\label{sec:conclusions}Conclusions}

The outcome of this work largely exceeded the initial expectations.
One goal was to test the performance of the simple EH algorithm (Eq.
\ref{eq: equacao simples}) and evaluate its suitability for further
development to create new algorithms for simulation of large and heterogeneous
systems. The second objective was testing of the new algorithm for
pseudo-diagonalization in realistic conditions. 

The simple EH approach (Eq. \ref{eq: equacao simples}) supplemented
with the nuclear-nuclear and nuclear-electronic term of Calzaferri
(Eq. \ref{eq: equacao correccoes electrostaticas}) performed remarkably
well. In the MD simulations the structural parameters (bond distances,
angles and torsions) compared very well with their experimental equivalents.
The EH parameters were optimized based on fitting of the potential
energy surface for each bond of NMA around the minimum. NMA provides
a similar, although not the same, chemical environment and the parameters
proved transferable. Transferability is a key concept in the development
of approximated computational methodologies allowing high quality
target data from smaller systems to be used in the parameterization.
The largest discrepancy to the experimental values was with the C\textsubscript{\textgreek{a}}-C\textsubscript{\textgreek{b}}
bond that was not included in the parameterization. These results
are very encouraging considering the simplicity of the electronic
Hamiltonian and the electrostatic corrections. There is a direct connection
between the extended Hückel and HF methods, meaning that suitable
approximations of the HF, or related correlated methods such as DFT,
will be possible. Our previous work on the computation of two-electron
integrals will be fundamental to derive computationally fast and accurate
approximations.

The algorithm for pseudo-diagonalization also performed very well.
Due to its iterative nature, it showed continuous improvement during
the MD simulations differing by \textasciitilde 0.18 kcal/mol relative
to the true diagonalization result. This is the first generation of
the algorithm and the main purpose was testing its usability. Subsequent
revisions will update the underlying optimization algorithms to faster
and more robust methods. The pseudo-diagonalization algorithms and
a detailed analysis of their performance will be the subject of a
dedicated publication.
\begin{acknowledgments}
P.E.M.L. wishes to thank M.M.G and J.D.N for support. 
\end{acknowledgments}

\bibliographystyle{apsrev4-1}
\bibliography{REFS_ARTIGOS_clean}

\end{document}